# Skyrmion Motion Driven by Oscillating Magnetic Field

Kyoung-Woong Moon[1], Duck-Ho Kim[2], Soong-Geun Je[2], Byong Sun Chun[1], Wondong Kim[1], Z.Q. Qiu[3], Sug-Bong Choe[2], and Chanyong Hwang[1*]

[1]*Center for Nanometrology, Korea Research Institute of Standards and Science, Daejeon 305-340, Republic of Korea.*

[2]*CSO and Department of Physics, Seoul National University, Seoul 151-742, Republic of Korea.*

[3]*Department of Physics, University of California at Berkeley, Berkeley, California 94720, USA.*

PACS number(s): 75.70.Kw, 75.78.Cd, 74.25.Ha

## Abstract

Magnetic skyrmion motion induced by an electric current has drawn much interest because of its application potential in next-generation magnetic memory devices. Recently, unidirectional skyrmion motion driven by an oscillating magnetic field was also demonstrated on large (20 µm) bubble domains with skyrmion topology. At smaller length scale which is more relevant to high-density memory devices, we here show by numerical simulation that a skyrmion of a few tens of nanometers could also be driven by high-frequency field oscillations but with the motion direction different from the tilted oscillating field direction. We found that high-frequency field for small size skyrmions could excite skyrmion resonant modes and that a combination of different modes would result in the final skyrmion motion with a helical trajectory. Because this helical motion depends on the frequency of the field, we can control both the speed and the direction of the skyrmion motion, which is a distinguishable characteristic compared with other methods.

---

[*]Corresponding author: cyhwang@kriss.re.kr

## I. INTRODUCTION

A magnetic skyrmion is a topological spin texture of which the spins wrap around a unit sphere [1-4]. A well-known example of skyrmion system is the chiral magnets in skyrmion crystal materials [4,5]. Recently, skyrmion control has been drawing much attention because of its use in noble spin devices as well as an unexpected underlying physics [6-13]. A precise control of the skyrmion motion relies on a full understanding of the skyrmion dynamics. The unidirectional motion of skyrmions, which can be realized by a magnetic field gradient [14], temperature gradient [11-13], and electric current flow in the sample [6-10], is expected to be useful for data transfer in memory and logic devices.

Recently, it has been demonstrated that a tilted alternating magnetic field can also induce such a unidirectional motion [15,16]. The direction of this unidirectional motion was towards the tilted field angle. Note that in this report, relatively large circular bubble domains (~10 μm in size) were used rather than small skyrmions (~10 nm in size) and the applied magnetic field had low frequency (< 1 kHz). In this paper, we expand our scope toward the small skyrmion beyond the large bubble and we expand the field frequency range up to GHz by employing micromagnetic simulations [17,18]. The field oscillation in the GHz frequency range can cause several resonant behaviors [19-24] of the skyrmion, such as a breathing mode [19,21] and a gyration mode [14,22,23]. We show that a combination of different resonance modes could not only drive the skyrmion motion but also control the skyrmion moving direction and speed by the field frequency and the skyrmion radius.

## II. SKYRMION MOTION: HELLICAL TRAJECTORY

We performed micromagnetic simulations using the OOMMF code [17] with the DMI extension module [18]. The material parameters were chosen as follows. The saturation magnetization $M_S$ was 580 kA m$^{-1}$, the exchange stiffness $A$ was 15 pJ m$^{-1}$, the uniaxial magnetic anisotropy $K_U$ was 0.8 MJ m$^{-3}$, the damping constant $\alpha$ was 0.3, and the DMI constant $D$ was –3.5 mJ m$^{-2}$. Most of these parameters have been used in Ref. [8] for stable skyrmion texture. We assumed a magnetic disk structure with a relatively large radius (150 nm) compared with the skyrmion radius (~15 nm) to observe almost free skyrmion motions not bounded by the disk edge [8,14,22,23]. The disk thickness $d$ was set to 0.4 nm. The unit cell was selected as 2 nm × 2 nm × 0.4 nm. The initial skyrmion state has its core at the center of the disk with a $+z$ directional polarity (up skyrmion). The spin texture in the central region of the disk is depicted in Fig. 1(a). The red and blue colors represent magnetizations in the $+z$ and $-z$ directions, respectively. The white color denotes the in-plane magnetization in the domain wall region with the black arrows indicating the in-plane magnetization direction. Because of the sufficient strength of the Dzyaloshinskii–Moriya interaction [25,26] (DMI) and the proper sign, the domain wall has radial outward magnetization [8]. The position of the skyrmion was determined by circular fitting of the domain wall positions. We defined the skyrmion radius $r$ as the distance from the skyrmion center to the domain wall position.

An oscillating external magnetic field was applied to verify the skyrmion motion. The field $\mathbf{H}$ is spatially uniform with a sinusoidal time ($t$) dependence, an amplitude of $H = 50$ mT, and a frequency of $f = 8$ GHz. The field is titled in the z-x plane with the tilting angle $\theta$ being $45°$ away from the z-axis. After the application of the field for 50 ns, the skyrmion was shifted ~40 nm from its initial position as shown in Fig. 1(a). Thus, we confirmed that an oscillatory magnetic field induces a motion of the skyrmion. Now, we will present the detailed characteristics of the oscillating field-induced skyrmion motion.

Figure 1(b) shows the detailed trajectory (red line) of the skyrmion center during the motion shown in Fig. 1(a). The skyrmion exhibits a helical motion, from which we can extract the steady motion (green arrow) with the velocity $\mathbf{V}$ defined as the average motion of the skyrmion over a long time period. It is notable that $\mathbf{V}$ is not in the $x$-direction, which is different from the magnetic bubble motion [15,16]. We defined this skyrmion motion direction with the azimuthal angle $\varphi$ from the $x$-axis. Figures 1(c) and 1(d) shows the skyrmion moving speed and direction as a function of the oscillatory field frequency at $H = 10$ mT amplitude and $\theta = 45°$. The skyrmion moving reaches maximum speed near 8 GHz, indicating a resonance behavior of the skyrmion under this condition. However, the moving direction does not exhibit any clear resonance behavior but rather an almost monotonic dependence on $f$.

Next, subtraction of the linear movement $\mathbf{V}t$ from the helical motion left the elliptic gyration with a clockwise rotation as shown by the blue line in Fig. 1(b). It is well known that the sign of the topological charge $q$ determines the rotation direction of the gyration and is typically determined by the polarity of the topological objects [2,14,22,23]. A prominent example is the gyrations of the vortex core. A magnetic vortex with $+z$ ($-z$) polarity has a nonzero positive (negative) $q$ thus should exhibit counterclockwise (clockwise) gyration [27,29].

Similarly, the sign of a skyrmion charge $q$ is determined by its polarity. According to previous studies [14,22,23] a skyrmion with $+z$ polarity should show counterclockwise gyration, but our results show the opposite case, i.e. clockwise gyration.

## II. SKYRMION OSCILLATION MODE

To study the characteristics of the skyrmion motion mentioned above, we decomposed the tilted oscillating field into perpendicular (along the $z$-axis) and in-plane (along the $x$-axis) components to check the magnetization oscillation modes driven by each of these two components.

The perpendicular field oscillation along the $z$-axis mainly causes the breathing mode of the skyrmion [19,21]. The breathing mode means that the oscillation of the skyrmion radius has a saturated amplitude $\Delta r$. The radius variation during the skyrmion motion is shown in the inset of Fig. 1(b). Since a perpendicular field does not break the radial symmetry, this field should only result in a radius change of the skyrmion without changing the core position and the radial symmetry. Figure 2(a) depicts the radius changing rates in term of $\Delta r f$. To obtain Fig. 2(a), a sinusoidal field was applied in the $z$ direction with an amplitude of $H = 5$ mT. The skyrmion radius changing rate $\Delta r f$ reaches the maximum value at $f \sim 8$ GHz which corresponds to the resonance frequency of the lowest skyrmion breathing mode [21]. This resonance behavior is also observed in Fig. 2(b), which shows a crossing of the breathing mode phase ($\delta_b$) at 90° near 8 GHz. It is notable that the resonance of the breathing mode corresponds to the maximum speed of the skyrmion motion [Fig. 1(c)], implying that the amplitude of the breathing mode should play an important role in determining the skyrmion moving speed.

The in-plane field generates the gyration motion [19,22,23,29,30]. We applied an oscillating in-plane fields in the $x$-axs with an amplitude of 5 mT. After turning on this field, the skyrmion eventually reaches a steady gyration without an obvious drifting. As seen in Fig 1(b), the trajectory of the gyration is an ellipse rather than a circle. Thus, the skyrmion gyration can be decomposed into two circular gyrations of opposite rotations. Figure 2(c) depicts how the two circular gyrations of opposite rotations generate an elliptic gyration. Two radii of the major ($R_L$) and minor ($R_S$) axes of the ellipse were measured, and then the radii of the two gyrations can be determined by $R_C = (R_L + R_S)/2$ and $R_{CC} = (R_L - R_S)/2$, where $R_C$ and $R_{CC}$ are the radius of the clockwise and counterclockwise gyrations, respectively. Figure 2(d) exhibits the values of $R_C$ and $R_{CC}$ as a function of the field frequency. The clockwise gyration has a larger amplitude than the counterclockwise gyration, resulting in a clockwise elliptic gyration of the skyrmion motion.

It was recently reported that a skyrmion confined in a circular disk has two kinds of gyration, clockwise and counterclockwise [22,23]. The resonance frequencies of these two circular gyrations are determined by three parameters; the gyrocoupling vector strength $\mathcal{G}$, the spring constant $\mathcal{K}$, and the skyrmion mass $\mathcal{M}$. In our model, we assumed a non-bounded free skyrmion ($\mathcal{K} = 0$). This condition results in two resonance angular frequencies, namely $\omega_+ \sim 0$ and $\omega_- \sim -\mathcal{G}/\mathcal{M}$, where $\omega_+$ is for counterclockwise gyration and $\omega_-$ is for clockwise gyration with the up-polarity skyrmion. This situation is similar to that in Fig. 2(d) because $R_{CC}$ shows a monotonous decrease with increasing frequency because of the zero resonance frequency. However, $R_C$ exhibits a resonance peak near 38 GHz, which corresponds to $-\mathcal{G}/\mathcal{M}$. Such resonance behavior is also shown in Fig. 2(b), which shows the phases of the clockwise and counterclockwise gyrations $\delta_C$ and $\delta_{CC}$, respectively. Note that $\omega_+$ originates from the massless motion of the skyrmion [22]. We can name the circular counterclockwise gyration "massless gyration." In contrast, $\omega_-$ depends on the skyrmion mass so that we can name the circular clockwise gyration "massive gyration."

### III. SKYRMION MOTION WITH CIRCULAR GYRATION

To separate the massive and massless gyrations in the skyrmion motion, we applied a rotating in-plane field rather than a uniaxial in-plane field because a rotating field should induce only one of its gyrations. We define the external field as $\mathbf{H} = (H_x, H_y, H_z) = (H_x^{osc}\sin\omega t, H_y^{osc}\cos\omega t, H_z^{osc}\sin\omega t)$, where $H_i$ is the field component along the $i$-axis, $H_i^{osc}$ is the field amplitude in the $i$ direction, and the angular frequency $\omega$ is $2\pi f$. Figure 3(a) presents the skyrmion trajectory up to 2 ns when $H_x^{osc} = H_y^{osc} = H_z^{osc} = 5$ mT and $f = 2$ GHz. In this case, we observe clockwise circular gyration with a steady motion of constant velocity $\mathbf{V}_C$. Changing the sign of $H_y^{osc}$, i.e. $H_y^{osc} = -5$ mT, the constant velocity $\mathbf{V}_{CC}$ and the circular gyration are retained but the gyration switches to counterclockwise because of the counterclockwise field rotation, as shown in Fig. 3(b).

These velocities depend strongly on the gyration radius as well as on $\Delta r$. Figure 3(c) presents the absolute value of the velocities as a function of the field frequency. $|\mathbf{V}_{CC}|$ shows a simple resonance behavior with a resonance peak near 8 GHz which is similar to $\Delta rf$ in Fig. 2(a). However, $|\mathbf{V}_C|$ exhibits a plateau in the frequency range of 8–30 GHz; thus we expect that the gyration radius is also involved in determining $|\mathbf{V}_C|$. The $\eta f \Delta r R_C$ and $\eta f \Delta r R_{CC}$ with $\eta = 0.12$ nm$^{-1}$ [Fig. 3(c)] show similar behaviors as $|\mathbf{V}_C|$ and $|\mathbf{V}_{CC}|$. However, there are notable discrepancies between $\eta f \Delta r R_C$ ($\eta f \Delta r R_{CC}$) and $|\mathbf{V}_C|$ ($|\mathbf{V}_{CC}|$) in the high-frequency regime. We think that the origin of this discrepancy [dashed lines in Fig. 3(c)] is related to the gyration radius because the discrepancy shows a similar trend to those of $R_C$ and $R_{CC}$ in Fig. 2(d).

The moving direction of the skyrmion also shows a dependence on the field rotation direction. We plotted the moving direction as a function of the field frequency in Fig. 3(d). The term $\varphi_C$ ($\varphi_{CC}$) is the angle between the direction of $\mathbf{V}_C$ ($\mathbf{V}_{CC}$) and the $x$-axis. We used the same definition of $\varphi$ as in Fig. 1(b). Note that the variation of $\varphi_C$ and $\varphi_{CC}$ is wider than that of $\varphi$. For example, the clockwise rotating field at $f = 1.5$ GHz induces a skyrmion motion along the $+x$ direction, whereas the clockwise rotating field at $f = 18$ GHz induces an almost $-y$-directional motion. This result shows that the frequency of the rotating magnetic field can control the direction of the skyrmion movement within a relatively wide angular range, which is not possible in previous methods such as the electric current flow [6-10] and the thermal or the field gradient [11-13] methods.

The skyrmion moving direction depends strongly on the phase difference between the breathing mode and the circular gyration mode, as shown in Fig. 3(d). This result indicates that the phase difference is the main factor in determining the skyrmion moving direction. The black closed (open) circles in Figs. 3(a) and 3(b) denote the positions where the skyrmion radius has its local maximum (minimum) values on the skyrmion trajectories. This result shows how the phase difference determines the skyrmion moving direction. In the clockwise (counterclockwise) rotation case, the skyrmion shifts towards where it has its maximum (minimum) radius [insets of Figs. 3(a) and 3(b)].

A superposition of two skyrmion motions of clockwise and counterclockwise gyrations gives raise exactly the final skyrmion motion driven by uniaxial oscillating field. This can be clearly seen in Fig. 1(c) that the $|\mathbf{V}_C + \mathbf{V}_{CC}|$ (blue line) describes exactly the $|\mathbf{V}|$. This result is also confirmed by comparing the moving direction of $\mathbf{V}_C + \mathbf{V}_{CC}$ [blue line in Fig. 1(d)] with $\varphi$.

## IV. SKYRMION MOTION AND BUBBLE MOTION

In Refs [15] and [16], magnetic bubble motions are demonstrated under a tilted oscillating magnetic field. Although the bubble motion shares some similarity as the skyrmion motion in the present study, their moving directions are different: the bubble moving direction is aligned to the tilting direction of the field, the skyrmion moving direction has a certain angle with respect to the tilting direction of the field. To explore the origin of this difference, we tested the size-dependent moving direction as shown in Fig. 4. The variation in material parameters resulted in the variation of the stabilized skyrmion radius. We selected and changed only one parameter in the simulation with a common oscillating field condition ($H$ = 10 mT, $f$ = 10 GHz, $\theta$ = 45°). The moving angles of the larger skyrmions were closer to 0° regardless of the detailed material parameters (Fig. 4). This means that the main source of the difference in moving direction was the skyrmion size.

## V. CONCLUSIONS

To summarize, we studied skyrmion motion within an oscillating magnetic field by micromagnetic simulations. A tilted uniaxial oscillating field produces the breathing and the gyration modes, and the combination of these two modes results in the skyrmion motion with a helical trajectory. This motion can be decomposed into clockwise and counterclockwise gyrations which are related to the massive and massless gyration modes, respectively. We studied these two skyrmion motions and found their resonance behaviors that determine the skyrmion speed and moving direction. The skyrmion motion induced by field oscillation has controllability of the direction of movement with frequency, which is a characteristic that is distinguishable from other methods.

**References**

[1]     T. H. R. Skyrme, *A Unified Field Theory of Mesons and Baryons*, Nucl. Phys. **31**, 556 (1962).

[2]     S. Heinze, K. von Bergmann, M. Menzel, J. Brede, A. Kubetzka, R. Wiesendanger, G. Bihlmayer, and S. Blügel, *Spontaneous Atomic-Scale Magnetic Skyrmion Lattice in Two Dimensions*, Nat. Phys. **7**, 713 (2011).

[3]     C. Pfleiderer, *Surfaces Get Hairy*, Nat. Phys. **7**, 673 (2011).

[4]     N. Nagaosa, and Y. Tokura, *Topological Properties and Dynamics of Magnetic Skyrmions*, Nat. Nanotech. **8**, 899 (2013).

[5]     S. Mühlbauer, B. Binz, F. Jonietz, C. Pfleiderer, A. Rosch, A. Neubauer, R. Georgii, and P. Böni, *Skyrmion Lattice in a Chiral Magnet*, Science **323**, 915 (2009).

[6]     F. Jonietz, S. Mühlbauer, C. Pfleiderer, A. Neubauer, W. Münzer, A. Bauer, T. Adams, R. Georgii, P. Böni, R. A. Duine, K. Everschor, M. Garst, and A. Rosch, *Spin Transfer Torques in MnSi at Ultralow Current Densities*, Science **330**, 1648 (2010).

[7]     R. Tomasello, E. Martinez, R. Zivieri, L. Torres, M. Carpentieri, and G. Finocchio, *A Strategy for the Design of Skyrmion Racetrack Memories*, Sci. Rep. **4**, 6784 (2014).

[8]     J. Sampaio, V. Cros, S. Rohart, A. Thiaville, and A. Fert, *Nucleation, Stability and Current-Induced Motion of Isolated Magnetic Skyrmions in Nanostructures*, Nat. Nanotech. **8**, 839 (2013).

[9]     N. Nagaosa and Y. Tokura, *Topological Properties and Dynamics of Magnetic Skyrmions*, Nat. Nanotech. **8**, 899 (2013).

[10]    J. Iwasaki, M. Mochizuki, and N. Nagaosa, *Universal Current-Velocity Relation of Skyrmion Motion in Chiral Magnets*, Nat. Commun. **4**, 1463 (2013).

[11]    M. Mochizuki, X. Z. Yu, S. Seki, N. Kanazawa, W. Koshibae, J. Zang, M. Mostovoy, Y. Tokura and N. Nagaosa, *Thermally Driven Ratchet Motion of a Skyrmion Microcrystal and Topological Magnon Hall Effect*, Nat. Mater. **13**, 241 (2014).

[12]  L. Kong, J. Zang, *Dynamics of an Insulating Skyrmion under a Temperature Gradient*, Phys. Rev. Lett. **111**, 067203 (2013).

[13]  K. Everschor, M. Garst, B. Binz, F. Jonietz, and S. Mühlbauer, *Rotating Skyrmion Lattices by Spin Torques and Field or Temperature Gradients*, Phys. Rev. B **86**, 054432 (2012).

[14]  C. Moutafis, S. Komineas, and J. A. C. Bland, *Dynamics and Switching Processes for Magnetic Bubbles in Nanoelements*, Phys. Rev. B **79**, 224429 (2009).

[15]  K.-W. Moon, D.-H. Kim, S.-C. Yoo, S.-G. Je, B. S. Chun, W. Kim, B.-C. Min, C. Hwang, and S.-B. Choe, *Magnetic Bubblecade Memory Based on Chiral Domain Walls*, Sci. Rep. **5**, 9166 (2015).

[16]  D. Petit, P. R. Seem, M. Tillette, R. Mansell, and R. P. Cowburn, *Two-Dimensional Control of Field-Driven Magnetic Bubble Movement Using Dzyaloshinskii–Moriya Interactions*, Appl. Phys. Lett. **106**, 022402 (2015).

[17]  Donahue, M. J. & Porter, D. OOMMF (NIST); http://math.nist.gov/oommf.

[18]  S. Rohart and A. Thiaville, *Skyrmion Confinement in Ultrathin Film Nanostructures in the Presence of Dzyaloshinskii-Moriya Interaction*, Phys. Rev. B **88**, 184422 (2013).

[19]  M. Mochizuki, *Spin-Wave Modes and Their Intense Excitation Effects in Skyrmion Crystals*, Phys. Rev. Lett. **108**, 017601 (2012).

[20]  N. Vukadinovic and F. Boust, *Three-Dimensional Micromagnetic Simulations of Multidomain Bubble-State Excitation Spectrum in Ferromagnetic Cylindrical Nanodots*, Phys. Rev. B **78**, 184411 (2008).

[21]  J.-V. Kim and F. Garcia-Sanchez, *Breathing Modes of Confined Skyrmions in Ultrathin Magnetic Dots*, Phys. Rev. B **90**, 064410 (2014).

[22]  I. Makhfudz, B. Krüger, and O. Tchernyshyov, *Inertia and Chiral Edge Modes of a Skyrmion Magnetic Bubble*, Phys. Rev. Lett. **109**, 217201 (2012).

[23]  K.-W. Moon, B. S. Chun, W. Kim, Z. Q. Qiu, and C. Hwang, *Control of Skyrmion Magnetic Bubble Gyration*, Phys. Rev. B **89**, 064413 (2014).

[24]    S.-Z. Lin, C. D. Batista, and A. Saxena, *Internal Modes of a Skyrmion in the Ferromagnetic State of Chiral Magnets*, Phys. Rev. B **89**, 024415 (2014).

[25]    I. Dzyaloshinskii, *A Thermodynamic Theory of 'Weak' Ferromagnetism of Antiferromagnetics*, J. Phys. Chem. Solids **4**, 241 (1958).

[26]    T. Moriya, *Anisotropic Superexchange Interaction and Weak Ferromagnetism*, Phys. Rev. **120**, 91 (1960).

[27]    B. Van Waeyenberge, A. Puzic, H. Stoll, K. W. Chou, T. Tyliszczak, R. Hertel, M. Fähnle, H. Brükl, K. Rott, G. Reiss, I. Neudecker, D. Weiss, C. H. Back, and G. Schütz, *Magnetic Vortex Core Reversal by Excitation with Short Bursts of an Alternating Field*, Nature (London) **444**, 461 (2006).

[28]    S.-B. Choe, Y. Acremann, A. Scholl, A. Bauer, A. Doran, J. Stöhr, and H. A. Padmore, *Vortex Core-Driven Magnetization Dynamics*, Science **304**, 420 (2004).

[29]    S.-K. Kim, K.-S. Lee, Y.-S. Yu, and Y.-S. Choi, *Reliable Low-Power Control of Ultrafast Vortex-Core Switching with the Selectivity in an Array of Vortex States by In-Plane Circular-Rotational Magnetic Fields and Spin-Polarized currents*, Appl. Phys. Lett. **92**, 022509 (2008).

[30]    A. A. Thiele, *Steady-State Motion of Magnetic Domains*, Phys. Rev. Lett. **30**, 230 (1973).

**Acknowledgments**

This work was supported by the Center for Advanced Meta-Materials(CAMM) funded by the Ministry of Science, ICT and Future Planning as part of the Global Frontier Project (CAMM-No. 2014063701, 2014063700) K.-W.M. has been supported by the TJ Park Science Fellowship of the POSCO TJ Park Foundation.

**Figure captions**

FIG. 1 (a) The uniaxial field oscillation and the magnetization states of the center of the disk. The field strength $H$ is 50 mT, the frequency $f$ is 8 GHz, and the tilting angle $\theta$ is 45°. The field is a sinusoidal function at time $t$. In magnetization images, the red region has perpendicular magnetization aligned along the $+z$ direction and the blue region corresponds to the $-z$ direction magnetization. The white color denotes in-plane ($x$–$y$ plane) magnetization and the black arrows represent in-plane magnetization directions. After application of the field during 50 ns, the skyrmion has shifted from its initial position. (b) The helical trajectory of the skyrmion center (red line) from 0 ns to 1 ns obtained in (a). The blue line represents the gyration motion obtained via subtraction of the steady motion $\mathbf{V}t$ from the helical motion. The inset shows the skyrmion radius variation during the motion. (c) Speeds of the steady motion $|\mathbf{V}|$ as a function of $f$ with $H = 10$ mT and $\theta = 45°$. (d) Directions of the skyrmion motion $\varphi$ with respect to the field frequency.

FIG. 2 (a) Rates of the skyrmion radius variation as a function of field frequency. To obtain this curve, the field was applied to the $z$ direction with a strength of 5 mT. This plot represents the resonance curve of the breathing mode of the skyrmion. (b) Phases of the breathing mode and the gyration modes. (c) Schematic diagram of the decomposition of the elliptic gyration into two circular gyrations having opposite rotating directions (clockwise, counterclockwise). (d) The resonance curves for the two gyration modes. The field was applied to the $x$ direction with a strength of 5 mT.

FIG. 3 (a) Trajectory of the skyrmion center with clockwise gyration (red line). (b) Trajectory with counterclockwise gyration (blue line). The magnetic field is ($H_x^{osc} \sin \omega t$, $H_y^{osc} \cos \omega t$, $H_z^{osc} \sin \omega t$), where $H_x^{osc} = H_z^{osc} = 5$ mT. $H_y^{osc} = 5$ mT for (a) and $H_y^{osc} = -5$ mT for (b). $f$ is 2 GHz. The black open (closed) circles represent the skyrmion center positions where the skyrmion radius has the local maximum (minimum) value. (c) Speeds of skyrmion motion as a function of field frequency. The red (blue) dashed line denotes the speed difference between $V_C$ ($V_{CC}$) and $\eta \Delta rf R_C$ ($\eta \Delta rf R_{CC}$). (d) Directions of skyrmion motion with respect to the field frequency and phase differences between the breathing and gyration modes.

FIG. 4 Moving directions of the small and large skyrmions. Skyrmions with different radii are obtained with material parameter variations. We selected only one parameter and changed that parameter to increase the skyrmion radius. For the larger skyrmion radiuses $d$ was increased to 2.6 nm, $M_S$ was increased to 667.4 kA m$^{-1}$, $A$ was reduced to 1.31 pJ m$^{-1}$, $K_U$ was reduced to 0.7258 MJ m$^{-3}$, and $D$ was increased to 3.746 mJ m$^{-2}$. Skyrmions with a radius larger than 30 nm were simulated in the disk with a diameter of 600 nm. A common oscillating field ($H = 10$ mT, $f = 10$ GHz, $\theta = 45°$) was applied.

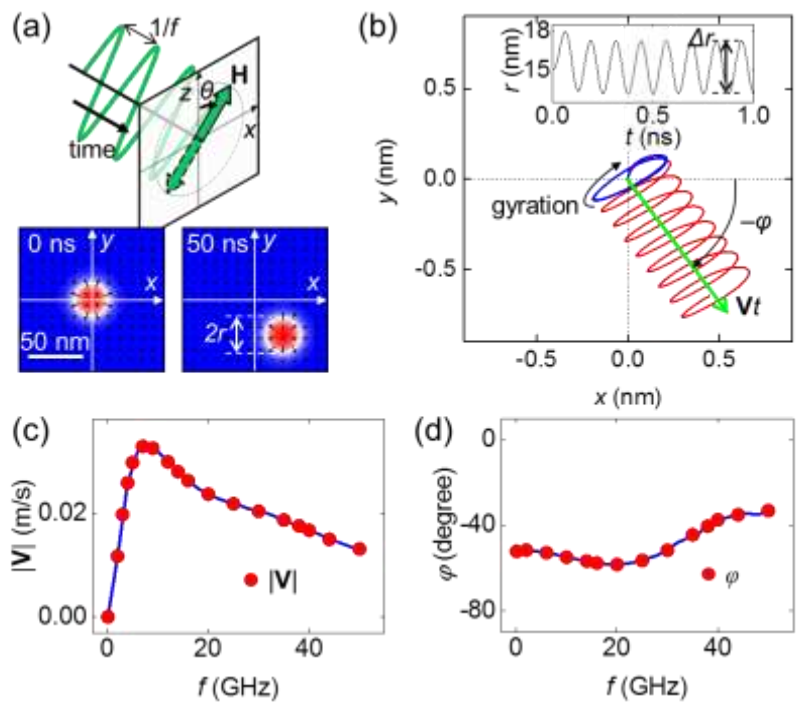

FIG. 1

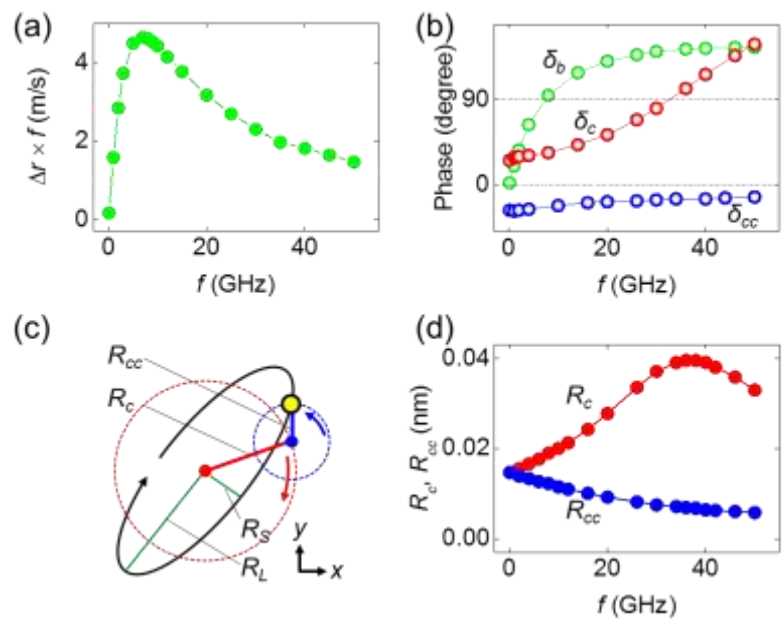

FIG. 2

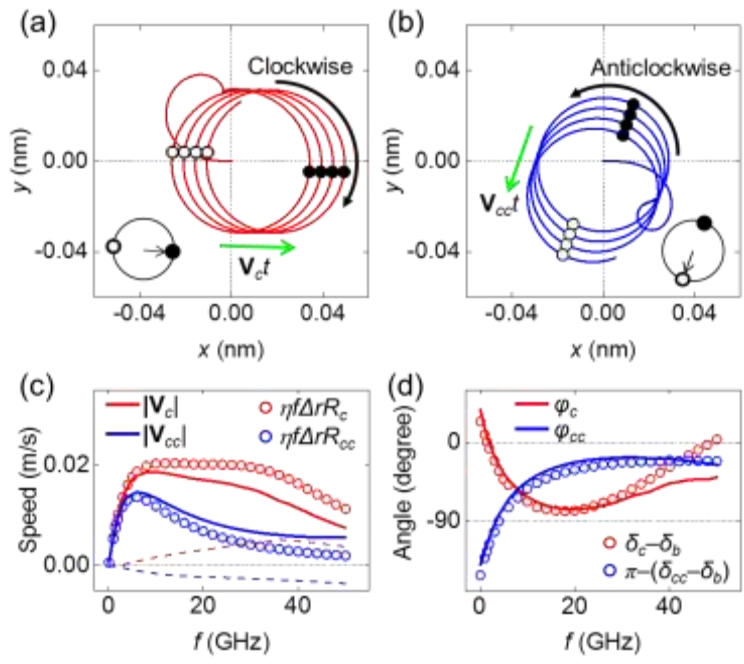

FIG. 3

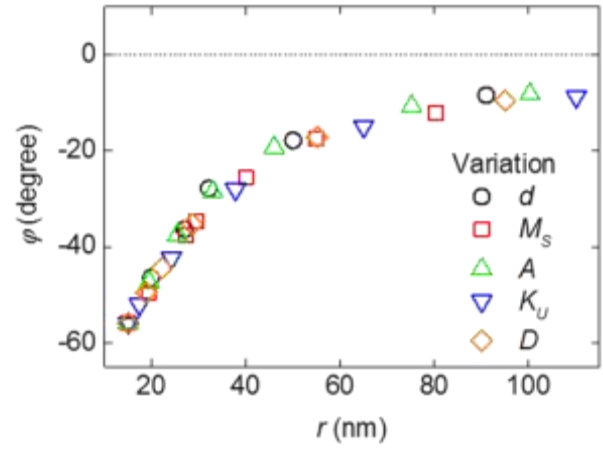

FIG. 4